# Suppression of superconductivity in layered $Bi_4O_4S_3$ by Ag doping

S. G. Tan,[1] P. Tong,[1] Y. Liu,[1] W. J. Lu,[1] L. J. Li,[1] B. C. Zhao,[1] Y. P. Sun,[1,2,*]

[1] Key Laboratory of Materials Physics, Institute of Solid State Physics, Chinese Academy of Sciences, Hefei 230031, People's Republic of China

[2] High Magnetic Field Laboratory, Chinese Academy of Sciences, Hefei 230031, People's Republic of China

**Abstract**

We report X-ray diffraction, magnetization and transport measurements for polycrystalline samples of the new layered superconductor $Bi_{4-x}Ag_xO_4S_3$ ($0 \leq x \leq 0.2$). The superconducting transition temperature ($T_C$) decreases gradually and finally suppressed for $x>0.10$. Accordingly, the resistivity changes from a metallic behavior for $x<0.1$ to a semiconductor-like behavior for $x>0.1$. The analysis of Seebeck coefficient shows there are two types of electron-like carriers dominate at different temperature regions, indicative of a multiband effect responsible for the transport properties. The suppression of superconductivity and the increased resistivity can be attributed to a shift of the Fermi level to the lower-energy side upon doping, which reduces the density of states at $E_F$. Further, our result indicates the superconductivity in the parent $Bi_4O_4S_3$ is intrinsic and the dopant Ag prefers to enter the $BiS_2$ layers, which may essentially modify the electronic structure.

*__Corresponding author__: Tel: +86-551-559-2757; Fax: +86-551-559-1434.
**E-mail:** ypsun@issp.ac.cn

## I. INTRODUCTION

To study the layered materials is one of the strategies for exploring new superconductors. The discoveries of superconductivity with high superconducting transition temperature ($T_C$) in cuprates and oxypnictides have given us good examples.[1,2] In those compounds, the $CuO_2$ layer or the $Fe_2An_2$ (An=P, As, Se, Te) layer plays a central role in determining the superconductivity. Very recently, Mizuguchi et al., reported the superconductivity in layered Bismuth oxy-sulfide $Bi_4O_4S_3$ with $T_C \sim 4.5$ K.[3] Structurally, $Bi_4O_4S_3$ is composed of a stacking of $Bi_4O_4(SO_4)$ spacer layers and two $BiS_2$ layers.[3] Band structure calculation indicates the Fermi level lies within the bands mainly originating from the Bi 6p orbitals within the $BiS_2$ layer. So the $BiS_2$ layer was believed to be a basic unit as the $CuO_2$ layer in cuprates or the $Fe_2An_2$ layer in oxypnictides.[3] Since the discovery of superconductivity, $Bi_4O_4S_3$ has attracted increasing attention because it could stimulate extensive investigations on the $BiS_2$-based layered compounds for exploring new high-$T_C$ superconductors.[4-9] The Hall coefficient measurements indicate an exotic multi-band behavior and the charge carriers are mainly electron-type,[4] which is consistent with our Seebeck coefficient measurement.[5] Our magnetization measurement implies that $Bi_4O_4S_3$ is a typical type-II superconductor.[5] Although the characterizations based on the polycrystalline samples with impurities, such as Bi and $Bi_2O_3$, indicated a bulk superconducting behavior,[6-7] it is questionable whether the superconductivity in $Bi_4O_4S_3$ is intrinsic because the almost pure sample synthesized by high pressure doesn't exhibit superconductivity down to 2K.[8] Further, chemical doping that usually serves as a powerful way to study the underlying physics of superconductivity, however, has not been reported up to date. Here we report our systematical structural, magnetic, electrical and thermal transport measurements for Ag doped $Bi_{4-x}Ag_xO_4S_3$ ($0 \leq x \leq 0.2$). The lattice shrinks gradually with Ag doping. The $T_C$ is reduced and finally suppressed for $x>0.1$. Accompanying the suppression of superconductivity, the resistivity is increased remarkably and behaves as a semiconductor for $x>0.1$, attributable to a shift of the Fermi energy ($E_F$) demonstrated by the analysis of Seebeck coefficient. For all samples, the Seebeck coefficient is linearly temperature dependent at both low- and high- temperature regions with different slopes, indicating a multiband behavior. Moreover, our results demonstrate that the superconductivity in $Bi_4O_4S_3$ is intrinsic and the dopant Ag enters the $BiS_2$ layer rather than the impurity Bi phase.

## II. EXPERIMENTAL DETAILS

The Ag, Bi, S and $Bi_2O_3$ powders with high purity in stoichiometric ratio were fully mixed and ground, and then pressed into pellets.[5] The pellets were sealed in an evacuated silica tube. The tube was heated at 450°C for 10 hours, and then at 510 °C for another 10 hours. The obtained pellets were ground again and the above treatment was repeated. The room-temperature crystal structure and lattice constants were determined by powder X-ray diffraction (XRD) (Philips X'pert PRO) with Cu $K_\alpha$ radiation. The electronic and thermal transport measurements were performed in a Quantum Design physical property measurement system (PPMS), and the magnetization measurement was performed on a superconducting quantum interference device

(SQUID) system.

## III. RESULT AND DISCUSSION

In Fig. 1(a), the XRD patterns are shown for the polycrystalline $Bi_{4-x}Ag_xO_4S_3$ ($0 \leq x \leq 0.2$) samples. The Bragg diffractions can be indexed using the tetragonal structure with the space group of $I4/mmm$.[3] The diffractions from the impurity Bi were also observed and one of them was marked by the asterisk. The diffractions from the $Bi_{4-x}Ag_xO_4S_3$ show a dependence on the Ag concentration $x$, while those from Bi don't. For instance, the (110) peak of $Bi_{4-x}Ag_xO_4S_3$ shifts clearly towards the higher angles as $x$ increases, as seen in Fig. 1(b). As shown in Fig. 1(c), however, the position and intensity of the Bi peak are unchangeable with $x$. This indicates that the dopant Ag indeed enters the $Bi_4O_4S_3$ lattice. The lattice parameters for $Bi_{4-x}Ag_xO_4S_3$ are obtained by a standard Rietveld refinement for the powder XRD patterns using the Rietica software.[10] As a result, both the lattice constant $a$ (=$b$) and $c$ decrease with increasing Ag content, as plotted in Fig. 2.

The temperature dependence of the resistivity $\rho(T)$ for $Bi_{4-x}Ag_xO_4S_3$ was measured by a standard four-probe method under a zero magnetic field. As displayed in Fig. 3(a), the metallic $\rho(T)$ shape at $x=0$ changes gradually to a semiconducting one when $x > 0.10$, accompanying a increase of the entire resistivity. Meanwhile, the superconducting transition temperature is reduced with increasing $x$ and non superconductivity can be found down to 2K for $x>0.1$ (Fig. 3(b)). The magnetic susceptibility $4\pi\chi(T)$ is displayed in Fig. 4. In good agreement with the resistivity result, the superconducting transition temperature decreases as $x$ increases. Simultaneously, the shielding volume fraction, determined as the value of $-4\pi\chi$ at 2 K, reduces from -0.45 emu/(cm$^3$ Oe) for $x = 0$ to $-3\times10^{-4}$ emu/(cm$^3$ Oe) for $x = 0.7$, above which the superconductivity, even if exists, lost its bulk characteristics. The superconducting transition temperatures defined by different ways will be discussed in the following text.

The Seebeck coefficient $S(T)$ is plotted in Fig. 5(a) for selected samples. The $S(T)$ is negative in the entire temperature range up to 320K, indicating the major carriers are electron-like for all samples. As shown in the inset of Fig. 5(a), the Seebeck coefficient at 320K, $S_{320K}$, increases as $x$ increases and a sharp enhancement happens at ~ $x=0.10$, consistent with the increasing resisitivity and the appearance of the semiconductor-like behavior for $x>0.1$ (Fig. 3). Generally, the $S(T)$ consists of three different parts, the diffusion term, the spin-dependent scattering term and the phonon-drag term due to electron-phonon interaction.[11] The spin-dependent contribution is probably very weak since there are no magnetic elements in $Bi_{4-x}Ag_xO_4S_3$. The phonon-drag effect gives ~ $T^3$ dependence for T«$\Theta_D$, and ~1/T for T ≥ $\Theta_D$ (where $\Theta_D$ is the Debye temperature), and thus a phonon-drag peak in $S(T)$ at ~ $\Theta_D/5$.[11] The $\Theta_D$ is 192K for the parent sample derived from heat capacity measurement.[6] However, the expected peak structure at ~ $\Theta_D/5 = 38K$ is absent and the high-temperature $S(T)$ doesn't obey the ~1/T expression, indicating the phonon contribution is considerably weak in the current samples. Usually, the diffusion term gives rise to a linearly temperature dependent $S(T)$ at low temperatures,[11] which is actually observed and shown in Fig.

5(b) (for clarity, only the data for $x=0$ and 0.2 are shown). Interestingly, at higher temperatures the $S$(T) exhibits another linear temperature dependence with a different slope for all samples investigated. For instance, the cases of $x=0$ and 0.2 were shown in the same figure. The high-temperature linear behavior of $S$(T) may indicate a different type of dominant charge diffusion contributing to the high-temperature $S$(T). The extrapolations of the linear $S$(T) portions at both low and high temperatures cross at some temperatures, marked as T*. The T* is found to be almost constant for $x<0.10$ within the error bars, but reduces rapidly above $x=0.10$, as shown in the inset of Fig. 5(b). Therefore, it indicates there exist two types of predominant electron-like charge carriers, one is active at low temperatures, the other at high temperatures.[12] As $x$ increases, the high-temperature charge carriers get more and more dominant, while the low-temperature one becomes weaker, shifting the T* to the lower temperatures. This behavior is possibly related with the $p_x$ and $p_y$ orbitals consisting the band structure near the $E_F$.[3] This result agrees in principle with the multiband feature of the Hall coefficient $R_H$(T) data.[4]

We are now focusing on the low-temperature linearly temperature dependent $S$(T) since the superconductivity occurs at the low temperatures. Given a single band mode with an energy-independent relaxation time, the slope dS/dT can be written as $S(T) = p^2 k_B^2 T /(3eE_F)$,[13] where $k_B$ is the Boltzman constant, $E_F$ the Fermi energy corresponds to band maximum or minimum. The $E_F$ can be estimated roughly using the slope of the low-temperature linear $S$(T), dS/dT. So, a linear fitting was applied to all data available below 50K. The value of dS/dT is found to increase rapidly with increasing $x$ when $x < 0.05$, and then saturates for $x \geq 1.0$ (see Fig. 6). It follows the evolution of the superconducting transition temperatures $T_C^{onset}$, defined as the onset of the dropping of the resistivity, and $T_C^0$, where the magnetic susceptibility -4πχ(T) drops, as shown in the same figure. Assuming a rigid band, the decreasing dS/dT indicates a shift of the $E_F$ to lower energy side. According to the theoretical calculation, for $x = 0$ the $E_F$ is just at the position of a sharp peak in the density of state (DOS) originating mainly from the Bi-6$p$ orbital within the $BiS_2$ layer.[3] Therefore, the shift of $E_F$ will reduce the DOS at $E_F$, $N(E_F)$, leading to the reduction of $T_C$. The reduced $N(E_F)$ can also account for the enhanced resistivity upon Ag doping. This result also indicates that the dopant Ag goes mainly to the $BiS_2$ layers, though Bi locates partially on the $Bi_2O_2$ layers. We note that the lattice effect on the electronic structure may be limited, because the observed lattice shrink is only ~ 0.6% along $c$ axis, and ~ 0.2% along $a(b)$ axis, by the doping level of $x = 0.2$. The external pressure that compresses the lattice suppresses the $T_C$, but also reduces the normal state resistivity,[9] contrary to the enhanced resistivity induced by Ag doping. Therefore, the suppression of superconductivity is mainly attributable to the modification of electronic structure, for instance, the shift of $E_F$ mentioned above, though detailed mechanisms need further efforts, particularly from the theoreticians.

## IV. CONCLUSION

In summary, polycrystalline $Bi_{4-x}Ag_xO_4S_3$ ($0 \leq x \leq 0.2$) samples were prepared, and physical properties investigated. The superconductivity is suppressed by $x=0.10$, above which the resistivity $\rho$(T) shows a semiconductor-like temperature dependence, in contrast with a metallic behavior for lower doping levels. The analysis of Seebeck coefficient indicates a multiband behavior and the shift of $E_F$ by Ag doping which may explain the suppression of $T_C$. All the results demonstrate that the dopant Ag enters the $Bi_4AgO_4S_3$ lattice and the superconductivity is rather intrinsic.

ACKNOWLEDGMENTS

This work was supported by the National Key Basic Research under contract No. 2011CBA00111 and the National Nature Science Foundation of China under contract Nos. 11274311, 11104279, 11174293 and 51102240, and Director's Fund of Hefei Institutes of Physical Science, Chinese Academy of Sciences.

**Figure captions:**

FIG. 1. (Color online) (a) The x-ray diffractions for $Bi_{4-x}Ag_xO_4S_3$ ($0 \leq x \leq 0.2$). One diffraction peak from impurity Bi was marked by the asterisk. The area of the (110) peak for $Bi_{4-x}Ag_xO_4S_3$ is enlarged in (b), and that of the Bi diffraction peak in (c).

FIG. 2. (Color online) The refined lattice constants for $Bi_{4-x}Ag_xO_4S_3$ ($0 \leq x \leq 0.2$).

FIG. 3. (Color online) (a) The resistivity for $Bi_{4-x}Ag_xO_4S_3$ ($0 \leq x \leq 0.2$) as a function of temperature. (b) An enlarged view for the low-temperature resistivity curves.

FIG. 4. (Color online) Temperature dependence of the magnetic susceptibility $4\pi\chi(T)$ for $Bi_{4-x}Ag_xO_4S_3$ ($0 \leq x \leq 0.07$) at $H=10$ Oe. Inset shows the details of the $4\pi\chi(T)$ curves around the $T_C$.

FIG. 5. (Color online) (a) Temperature dependence of the Seebeck coefficient for $Bi_{4-x}Ag_xO_4S_3$ ($0 \leq x \leq 0.2$). Inset shows the Seebeck coefficient at 320K, $S_{320K}$. (b) The extrapolations (dotted lines) of both low-temperature and high-temperature linear $S(T)$ curves for $x=0$ and 0.20. The crossing temperature T* for the two linear $S(T)$ parts is plotted in the inset as a function of $x$.

FIG. 6. (Color online) The slope dS/dT of the linearly temperature dependent $S(T)$ below 50K, and the superconducting transition temperatures, $T_C^{onset}$ and $T_C^0$ (see text for definitions) as a function of Ag composition. The hatched bar indicates the superconducting-nonsuperconducting boundary.

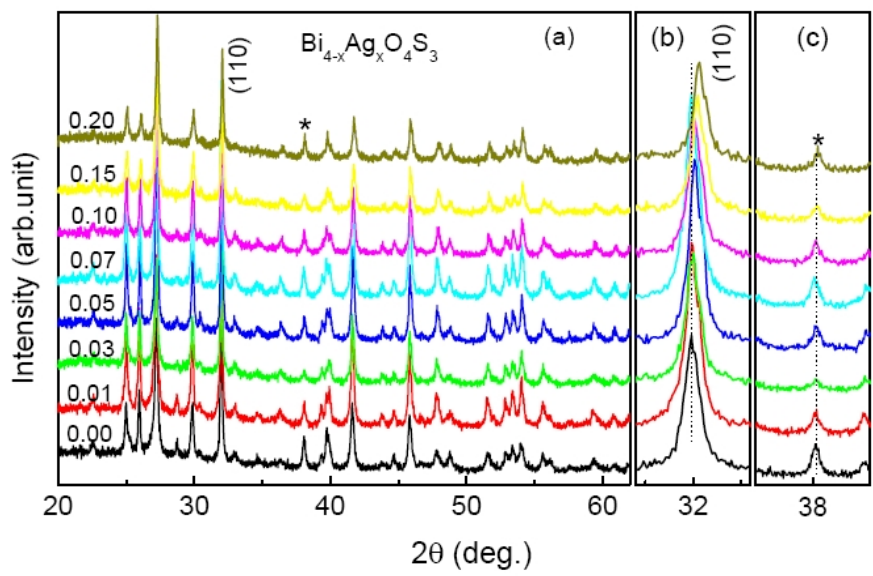

FIG. 1. Tan et al.,

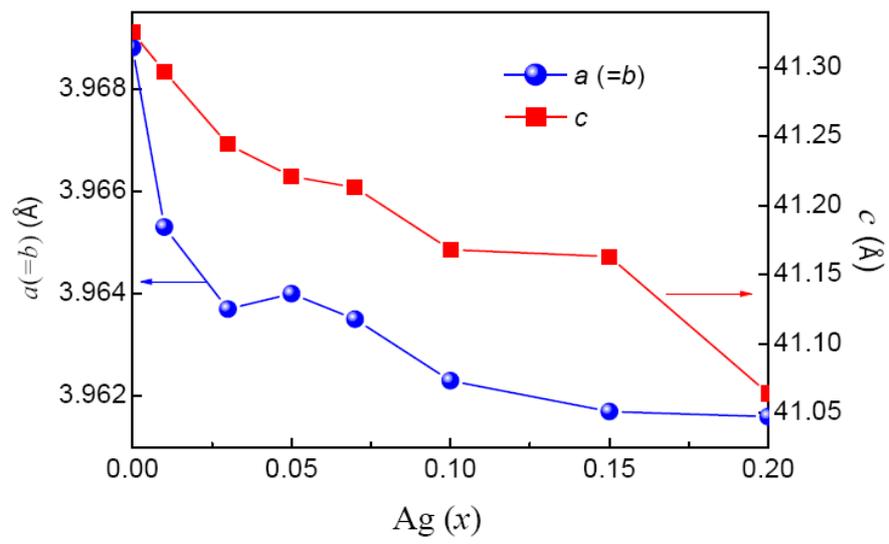

FIG. 2. Tan et al.,

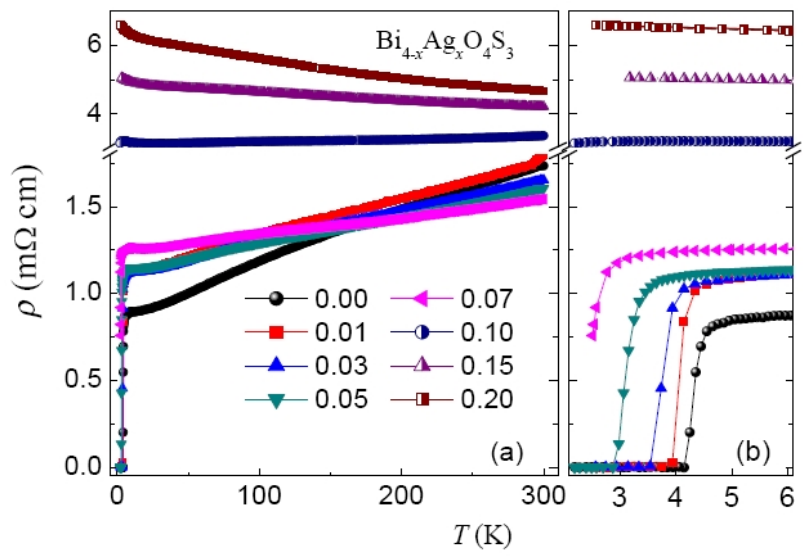

FIG. 3. Tan et al.,

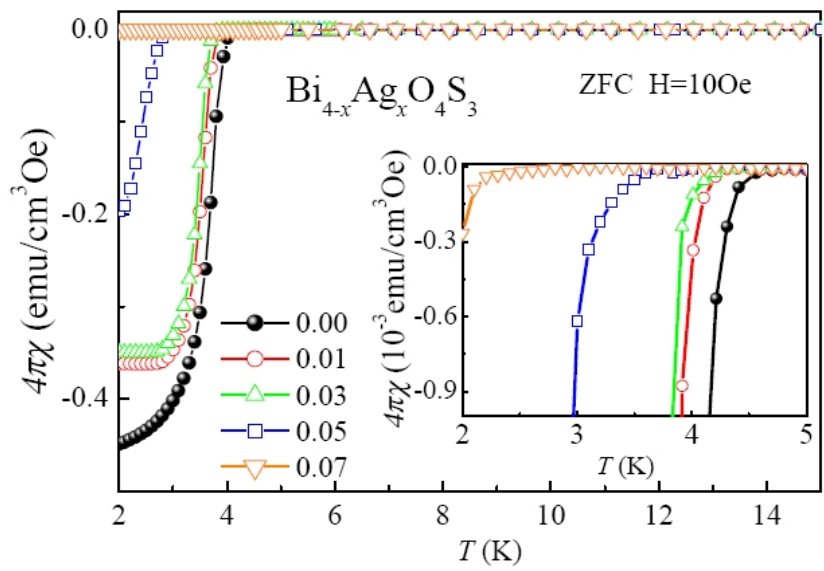

FIG. 4. Tan et al.,

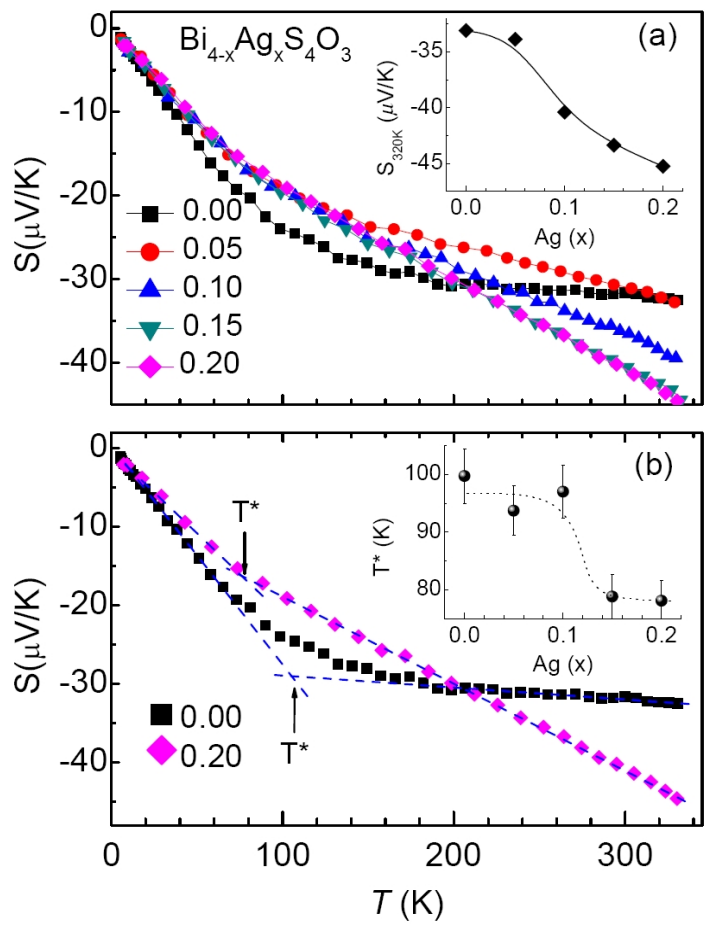

FIG. 5. Tan et al.,

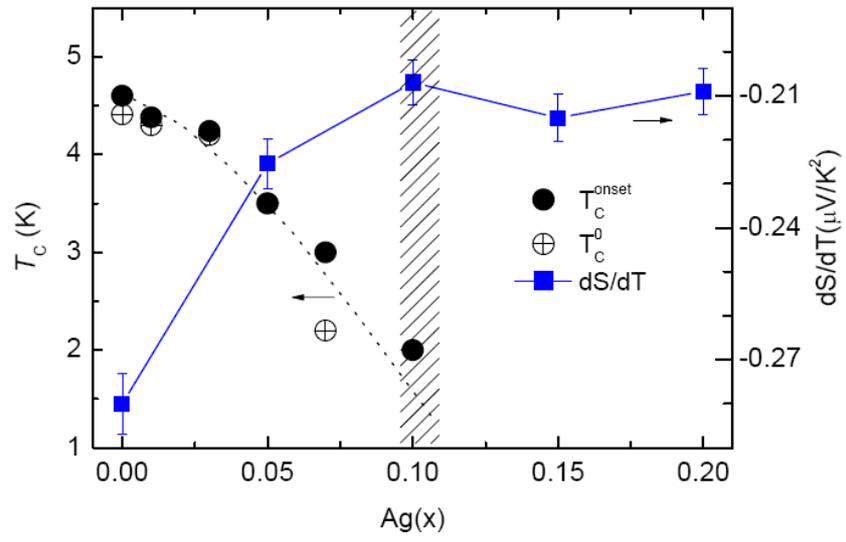

FIG. 6. Tan et al.,